\documentclass[draftclsnofoot,onecolumn]{IEEEtran}
\IEEEoverridecommandlockouts
\usepackage{subfigure} 
\usepackage{graphicx}

\usepackage{amsmath,graphicx,amssymb,mathtools,bm}
\usepackage{subfigure}
\usepackage{hyperref}
\usepackage{cite}
\usepackage{amsmath,amssymb,amsfonts}
\usepackage{algorithmic}
\usepackage{graphicx}
\usepackage{textcomp}
\usepackage{xcolor}
\usepackage{verbatim}  
\usepackage{graphicx}  
\usepackage{bm}  
\usepackage{mathrsfs} 
\usepackage{algorithm} 
\usepackage{algorithmic} 
\usepackage{booktabs}
\usepackage{textcomp}  
\usepackage{multirow}  
\usepackage{lettrine}   
\usepackage{graphicx}  
\usepackage{color}  
\usepackage{amsmath}
\usepackage{amssymb}
\usepackage{stfloats}

\def\BibTeX{{\rm B\kern-.05em{\sc i\kern-.025em b}\kern-.08em
    T\kern-.1667em\lower.7ex\hbox{E}\kern-.125emX}}
\begin{document}
	\newcommand{\tabincell}[2]{\begin{tabular}{@{}#1@{}}#2\end{tabular}} 

\title{Fast Compressive Channel Estimation for MmWave MIMO Hybrid Beamforming Systems
}

\author{Songjie Yang, Chenfei Xie, Dongli Wang and
        Zhongpei Zhang,~\IEEEmembership{Member,~IEEE},

\thanks{This work was supported in part by the National Key Research and Development Program of China under Grant 2020YFB1806805. (\textit{Corresponding author:
Zhongpei~Zhang}.)
}

\thanks{All authors are with the National Key Laboratory of Science and Technology on Communications, University of Electronic Science and Technology of China, Chengdu 611731, China. (e-mail:
	yangsongjie@std.uestc.edu.cn; 201911220505@std.uestc.edu.cn;202021220519@std.uestc.edu.cn;
	zhangzp@uestc.edu.cn).}
}
\maketitle

\begin{abstract}
Given the high degree of computational complexity of the channel estimation technique based on the conventional one-dimensional (1-D) compressive sensing (CS) framework employed in the hybrid beamforming architecture, this study proposes two low-complexity channel estimation strategies. One is two-stage CS, which exploits row-group sparsity to estimate angle-of-arrival (AoA) first and uses the conventional 1-D CS method to obtain angle-of-departure (AoD). The other is two-dimensional (2-D) CS, which utilizes a 2-D dictionary to reconstruct the 2-D sparse signal. To conduct a meaningful comparison of the three CS frameworks, i.e., 1-D, two-stage and 2-D CS, the orthogonal match pursuit (OMP) algorithm is employed as the basic algorithm and is expanded to two variants for the proposed frameworks. Analysis and simulations demonstrate that when the 1-D CS method is compared, two-stage CS has somewhat lower performance but significantly lower computational complexity, while 2-D CS is not only the same as 1-D CS in terms of performance but also slightly lower in computational complexity than two-stage CS.

\end{abstract}
\section{Introduction}
\lettrine[lines=2]{M}assive multiple-input multiple-output (MIMO) is a potential millimeter-wave (mmWave) communication technique that offers higher communication performance \cite{mmwave1,mmwave3}. However, fully-digital antenna arrays will incur substantial hardware costs and energy consumption, which is one of the primary difficulties with massive MIMO. As a result, the hybrid beamforming architecture has been widely used to overcome the problem of high cost and power consumption of mmWave mixed-signal hardware \cite{HB1,HB2,HB3}. The challenge with the hybrid beamforming architecture, however, is obtaining reliable channel state information owing to the large number of antennas and several radio frequency (RF) chains employed.

Due to the high training overhead required for low-frequency channel estimate techniques such as the least squares (LS) algorithm, they are no longer fit for purpose to hybrid architecture.
Fortunately, the angular sparsity property of mmWave channels is benificial for channel estimation with low training overhead. Up to date, a variety of sparse channel estimation methods based on compressive sensing (CS), i.e., angle-of-arrival (AoA) and angle-of-departure (AoD) estimation, have been presented for diverse contexts\cite{CS1,CS2,CS3,CS4,CS5,CS6,CS7,multi2,omp,LASSO,sbl1,sbl2}. For instance, the authors in \cite{LASSO} presented an optimal-tuned weighted LASSO for channel estimation in massive MIMO communications. Additionally, sparse Bayesian learning (SBL) based methods for channel estimation were investigated in \cite{sbl1,sbl2}, which demonstrated great performance but required a high level of computational complexity. Recently, in \cite{off1,off2}, the off-grid theory utilized to tackle the CS mismatch problem \cite{mismatch} was employed to further improve channel estimation performance, although at the expense of increased computational complexity. On the other hand, channel estimation with the hybrid beamforming architecture has been vestigated in \cite{hpce1,hpce2,hpce3}

However, almost all existing sparse channel estimation algorithms take into account the one-dimensional (1-D) CS framework, which stacks the matrix into a long vector via the Kronecker product but ignores the dimension feature. When confronted with multidimensional estimation problems, such as the two-dimensional issue of AoA/AoD estimation, this CS framework results in significantly increased computational complexity, as well as memory usage. Furthermore, the large-scale antenna array itself carries with a high computational complexity that cannot be ignored. Since the high computational complexity will preclude the real system from being implemented, it is of paramount significance to develop novel low-complexity channel estimation schemes. This motivates us to explore computationally efficient channel estimation techniques.

In this study, we propose two low-complexity channel estimation frameworks based on the CS technique. The first proposed framework estimates AoA first via row-group sparsity and, subsequently, AoD with the conventional 1-D CS framework. Further, we convert the conventional 1-D CS framework into the 2-D CS framework by developing a two-dimensional (2-D) dictionary. The two proposed methods significantly lower the computational complexity when compared to the conventional 1-D CS framework.
 
 \emph{Notations:} We use the following notations throughout this study. $\mathbb{C}^{x \times y}$ represents the complex-value matrices with the space of $ x \times y $. $(\cdot)^T$, $(\cdot)^*$ and $(\cdot)^H$ denote transpose, conjugate and conjugate transpose, respectively. $\Vert \cdot\Vert_0$, $\Vert \cdot\Vert_2$, $\Vert \cdot\Vert_F$ are $\ell_0$ norm, $\ell_2$ norm and Frobenius norm respectively.  {\rm tr($\cdot$)} is the trace of the matrix, $\left\langle\cdot,\cdot\right\rangle$ denotes the inner product. $a^n$ is the $n$-th element of vector $\mathbf{a}$.
  Further, $\rm vec(\cdot)$ and $\rm decec(\cdot)$ represent the vectorization and devectorization operations, respectively.
  $\mathbf{I}_M$ denotes the $M$-by-$M$ identity matrix. Finally, $\otimes$ denotes the Kronecker product and $\odot$ is the Khatri-Rao product.
\section{System and Channel Model}

We consider a widely used mmWave massive MIMO communication system with the fully-connected hybrid beamforming architecture, as shown in \figurename{1}, where the transmitter (TX) and the receiver (RX) are equipped with $N_t$ and $N_r$ antennas respectively, and both of them have $N_{RF}$ RF chains. 
\begin{figure*}[htbp]\label{structure1}
	\centering
	\includegraphics[width=17cm,height=6.6cm]{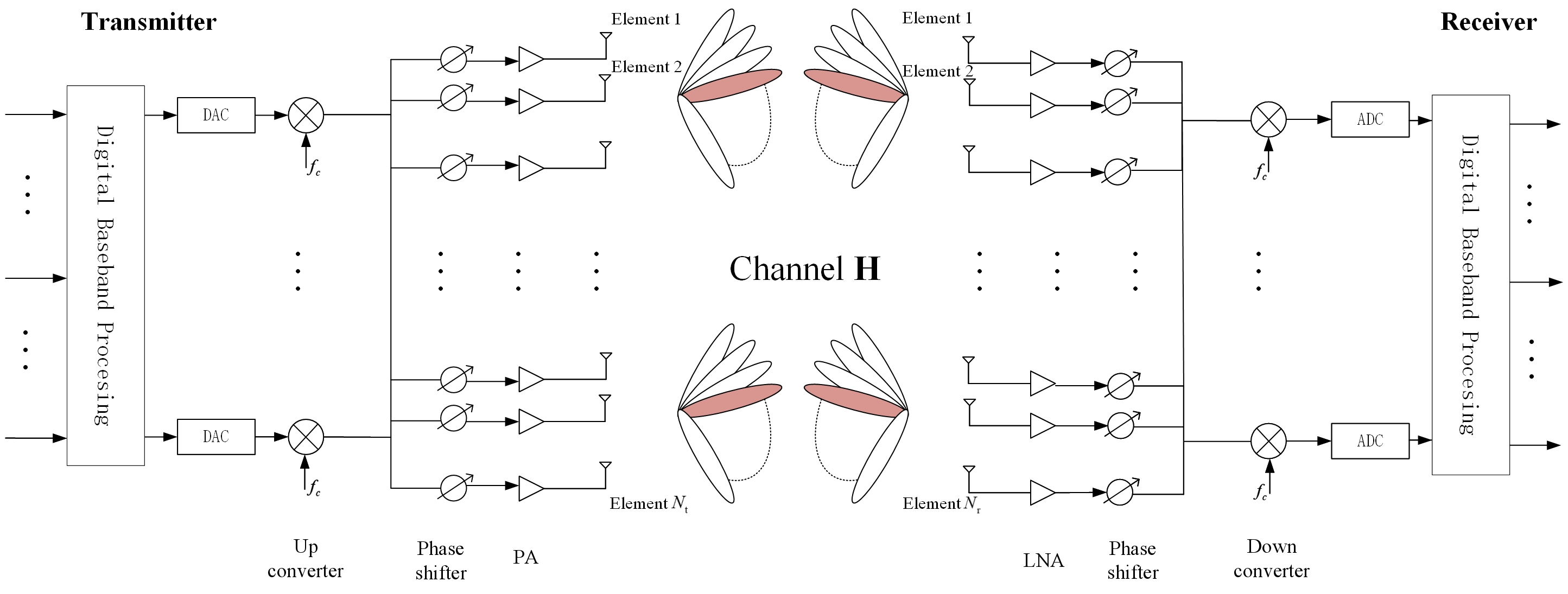}
	\centering
	\caption{A communication system of the fully-connected hybrid beamforming architecture with $N_t$ and $N_r$ antennas. }
\end{figure*}
\subsection{Signal Model}
Postulate that the TX sends $N_X$ pilots, and the RX spends $Q$ time slots receiving the signal $\mathbf{y}_k\in\mathbb{C}^{N_Y\times1}$, where $N_Y=QN_{RF}$. Denoting the $k$-th pilot at the $q$-th time slot as $\mathbf{y}_{k, q}=\mathbf{w}_{q}^{H}\left(\mathbf{H f}_{q} \mathbf{x}_{k, q}+\mathbf{n}_{k, q}\right)$, the received pilots can be collected and expressed as
\begin{equation}
 \mathbf{y}_{k}=\mathbf{W}^{H}\mathbf{HF}\mathbf{x}_{k}+\mathbf{W}^{H}\mathbf{n}_{k},
\end{equation}
where $\mathbf{x}_{k}= \left[\mathbf{x}_{k, 1}^{T}, \mathbf{x}_{k, 2}^{T}, \cdots, \mathbf{x}_{k, Q}^{T}\right]^{T}$, $\mathbf{y}_{k}=\left[\mathbf{y}_{k, 1}^{T}, \mathbf{y}_{k, 2}^{T}, \cdots, \mathbf{y}_{k, Q}^{T}\right]^{T}$, $\mathbf{H}$ is the channel matrix between the TX and the RX, $ \mathbf{n}_{k}=\left[\mathbf{n}_{k, 1}^{T}, \mathbf{n}_{k, 2}^{T}, \cdots, \mathbf{n}_{k, Q}^{T}\right]^{T}$ and $\mathbf{n}_{k,q}$ is the independent and identically distributed (i.i.d) additive white Gaussian noise following the distribution $\mathcal{C} \mathcal{N}\left(0, \sigma_n^{2} \mathbf{I}_{N_{\mathrm{RF}}}\right)$. In addition, 
$\mathbf{F}=\left[\mathbf{f}_{1}, \mathbf{f}_{2}, \cdots, \mathbf{f}_{Q}\right] , \mathbf{W}=\left[\mathbf{w}_{1}, \mathbf{w}_{2}, \cdots, \mathbf{w}_{Q}\right]$ and 
$\mathbf{f}_{q}$ and $\mathbf{w}_{q}$ are the hybrid precoder and combiner respectively. 

Then the received signal by collecting $N_X$ transmitted pilots can be given by 
\begin{equation}
	\mathbf{Y}=\mathbf{W}^H\mathbf{HFX}+\mathbf{W}^H\mathbf{N},
\end{equation}
where $\mathbf{X}=[\mathbf{x}_1,\mathbf{x}_2,\cdots,\mathbf{x}_{N_X}]$, $\mathbf{Y}=[\mathbf{y}_1,\mathbf{y}_2,\cdots,\mathbf{y}_{N_X}]$ and $\mathbf{N}=[\mathbf{n}_1,\mathbf{n}_2,\cdots,\mathbf{n}_{N_X}]$. In addtion, we postulate the transmitted pilot signal $\mathbf{X}$ satisfy orthogonality, i.e., $\mathbf{X X}^{H}=\sigma_p^2 \mathbf{I}_{N_{\mathrm{X}}}$, where $\sigma_p^2$ is the transmit power and is set to 1 for study. The hybrid precoder $\mathbf{F}$ and combiner $\mathbf{W}$ can be designed as simplified forms, where $\mathbf{F}=\left[\mathbf{I}_{N_{\mathrm{X}}}, \mathbf{0}_{N_{\mathrm{X}} \times\left(N_{t}-N_{\mathrm{X}}\right)}\right]^{T} \in \mathbb{C}^{N_{t} \times N_{\mathrm{X}}}, \mathbf{W}=$ $\left[\mathbf{I}_{N_{\mathrm{Y}}}, \mathbf{0}_{N_{\mathrm{Y}} \times\left(N_{r}-N_{\mathrm{Y}}\right)}\right]^{T} \in \mathbb{C}^{N_{r} \times N_{\mathrm{Y}}}$.
\subsection{Channel Model}
Next, we specify the physical channel which can characterize
the geometrical structure and limited scattering nature of mmWave
channels. The channel matrix is written as
\begin{equation}\label{H}
	\mathbf{H}=\sqrt{\frac{N_tN_r}{L}}\sum_{l=1}^{L}z_l\mathbf{a}_r(\phi_l)\mathbf{a}_t^H(\theta_l),
\end{equation}
where $L$ is the number of paths, $z_l$ denotes the complex gain of the $l$-th path, $\mathbf{a}_r(\phi_l)$ and $\mathbf{a}_t(\theta_l)$ are array response vectors expressed as
\begin{equation}
	\begin{aligned}
			&\mathbf{a}_{t}\left(\theta_{l}\right)=\sqrt{1/N_t}\left[1, e^{j 2 \pi d \sin \theta_{l} / \lambda}, \cdots, e^{j 2 \pi d\left(N_{t}-1\right) \sin \theta_{l} / \lambda}\right]^{T}, \\
		&\mathbf{a}_{r}\left(\phi_{l}\right)=\sqrt{1/N_r}\left[1, e^{j 2 \pi d \sin \phi_{l} / \lambda}, \cdots, e^{j 2 \pi d\left(N_{r}-1\right) \sin \phi_{l} / \lambda}\right]^{T},
	\end{aligned}
\end{equation}
where $\lambda$ is the antenna wavelength and $d$ is the antenna element spacing.

The compact form of Eqn. (\ref{H}) can be written as
\begin{equation}
	\mathbf{H}=\mathbf{A}_{\mathrm{R}}{\rm diag}(\bm{z})\mathbf{A}_{\mathrm{T}}^H,
\end{equation} 
where $\bm{z}=\sqrt{N_tN_r/L} \ [z_1,z_2,\cdots,z_L]^T$, 
\begin{equation}
	 \mathbf{A}_{\mathrm{T}}=[\mathbf{a}_t(\theta_1),\mathbf{a}_t(\theta_2),\cdots,\mathbf{a}_t(\theta_L)]
\end{equation}
and
\begin{equation}
	\mathbf{A}_{\mathrm{R}}=[\mathbf{a}_r(\phi_1),\mathbf{a}_r(\phi_2),\cdots,\mathbf{a}_r(\phi_L)].
\end{equation}
\section{Three CS Frameworks-based MmWave Channel Estimation}\label{PF}
In this section, the conventional channel estimation method based on 1-D CS framework is first reviewed. Then we present two novel channel estimation frameworks on the basis of CS.
\subsection{Conventional 1-D CS}

To exploit the property of the sparse channel to form a CS problem, the matrix $\mathbf{Y}$ after cancelling the effect of the training sequences is vectorized as 
\begin{equation}\label{1D}
\begin{aligned}
	\tilde{\mathbf{y}}&= \operatorname{vec}\left(\mathbf{W}^{H} \mathbf{H F}\right)+\operatorname{vec}(\mathbf{W}^H\mathbf{N}\mathbf{X}^H) \\
	&=\left(\mathbf{F}^{T} \otimes \mathbf{W}^{H}\right) \operatorname{vec}(\mathbf{H})+\tilde{\mathbf{n}}\\
	&=\left(\mathbf{F}^{T} \otimes \mathbf{W}^{H}\right)\left(\mathbf{A}_{\mathrm{T}}^{*} \odot \mathbf{A}_{\mathrm{R}}\right) \bm{z}+\tilde{\mathbf{n}},
\end{aligned}
\end{equation}
where $\mathbf{A}_{\mathrm{T}}^{*} \odot \mathbf{A}_{\mathrm{R}}$ is an  $N_{t} N_{r} \times L$ matrix in which each column has the form $\left(\mathbf{a}_{t}^{*}\left(\theta_{l}\right) \otimes \mathbf{a}_{r}\left(\phi_{l}\right)\right), l=1,2, \ldots, L$, that is, the $l$-th column denotes the Kronecker product of the TX and RX array response vectors in light of the AoA/AoD of the $l$-th path of the channel.

By using the 1-D dictionary matrix $\bar{\mathbf{A}}=\bar{\mathbf{A}}_{\rm T}^*\otimes\bar{\mathbf{A}}_{\rm R}\in\mathbb{C}^{N_tN_r\times N^2}$ associated with $N$ AoAs/AoDs, Eqn. (\ref{1D}) can be approximated as
\begin{equation}\label{1DCS}
	\tilde{\mathbf{y}}=\left(\mathbf{F}^{T} \otimes \mathbf{W}^{H}\right)\bar{\mathbf{A}} \tilde{\bm{z}}+\tilde{\mathbf{n}},
\end{equation}
where $\bar{\mathbf{A}}_{\rm T}=[\mathbf{a}(\theta_1),\mathbf{a}(\theta_2),\cdots,\mathbf{a}(\theta_N)]\in\mathbb{C}^{N_t\times N}$ and $\bar{\mathbf{A}}_{\rm R}=[\mathbf{a}(\phi_1),\mathbf{a}(\phi_2),\cdots,\mathbf{a}(\phi_N)]\in\mathbb{C}^{N_r\times N}$
 . The AoDs $\{\theta_i\}_{i=1}^N$ and AoAs $\{\phi_i\}_{i=1}^N$ are taken from the Discrete Fourier Transformation (DFT) bin, i.e., $\sin(\theta_i/\phi_i)\in\{ -\frac{1}{2},-\frac{1}{2}+\frac{1}{N},\cdots,\frac{1}{2}-\frac{1}{N}\}$. 
Hence, the channel estimation problem on the basis of 1-D CS framework is given by
\begin{equation}\label{1d}
	\begin{aligned}
		&	\underset{\mathbf{z}}{\rm arg \ min} \ \Vert \mathbf{z}\Vert_0\\
		{\rm s.t.} \ &\Vert \tilde{\mathbf{y}}-
		\tilde{\mathbf{A}}\mathbf{z} \Vert_2 <\varepsilon,
	\end{aligned}
\end{equation}
where $\tilde{\mathbf{A}}=\left(\mathbf{F}^{T} \otimes \mathbf{W}^{H}\right)\bar{\mathbf{A}}\in\mathbb{C}^{N_XN_Y\times N^2}$, $\tilde{\mathbf{y}}$ is the measurement vector and $\varepsilon$ is the precision parameter. 

According to Eqn. (\ref{1d}), a variety of compressive sensing algorithms are capable of recovering the sparse vector $\mathbf{z}$. However, this 1-D CS framework will lead to a huge amount of computational complexity due to the long length of 1-D vector generated by the vectorization. This can be addressed by the frameworks we present next.
\subsection{Two-Stage CS}
The 2-D received signal after cancelling the training sequences can be expressed as
\begin{equation}
\tilde{\mathbf{Y}}=\tilde{\mathbf{A}}_{\mathrm{R}} \operatorname{diag}(\bm{z}) \tilde{\mathbf{A}}_{\mathrm{T}}^{H}+\tilde{\mathbf{N}},
\end{equation}
where $\tilde{\mathbf{A}}_{\mathrm{R}}=\mathbf{W}^{H} \bar{\mathbf{A}}_{\mathrm{R}}\in\mathbb{C}^{N_Y\times N} , \tilde{\mathbf{A}}_{\mathrm{T}}^H=\bar{\mathbf{A}}_{\mathrm{T}}^{H} \mathbf{F}\in\mathbb{C}^{N\times N_X}$ are respectively the new manifold matrices and $\tilde{\mathbf{N}}=\mathbf{W}^{H} \mathbf{N X}^{H}$ is noise matrix.

The key idea of the two-stage CS framework is to estimate AoAs first and then AoDs, or vice versa. Indeed, there exists a favorable property in the first estimation stage, i.e., group sparsity. In short, row-/column-group sparisty refers to that each column/row of the matrix to be estimated has the same non-zero element index.

In the AoAs estimation stage, the CS problem with row-group sparsity can be given by
\begin{equation}\label{2st}
	\begin{aligned}
		&\underset{\mathbf{Z}_{\rm T}}{\rm arg \ min} 	\ \Vert \mathbf{Z}_{\rm T}\Vert_0\\
		{\rm s.t.} \ &\Vert \tilde{\mathbf{Y}}-
		\tilde{\mathbf{A}}_{\rm R}\mathbf{Z}_{\rm T} \Vert_F <\varepsilon,\\
		{\rm supp}&({\mathbf{z}_{{\rm T}_1}})=\cdots={\rm supp}({\mathbf{z}_{{\rm T}_{N_X}}}),
	\end{aligned}
\end{equation}
where $\mathbf{Z}_{\rm T}=[\mathbf{z}_{{\rm T}_1},\mathbf{z}_{{\rm T}_2},\cdots,\mathbf{z}_{{\rm T}_{N_X}}]$ and ${\rm supp}(\mathbf{v})$ represents the indexes of non-zero elements of an arbitrary vector $\mathbf{v}$. Denoting $\tilde{\mathbf{Z}}_{\rm T}\in\mathbb{C}^{L\times N_X}$ as the estimated group signal after removing all zero rows, then AoDs can be attained by the typical 1-D CS method, which is described as
\begin{equation}
	\begin{aligned}
		&	\underset{\mathbf{z}}{\rm arg \ min} \ \Vert \mathbf{z}\Vert_0\\
		{\rm s.t.} \ &\Vert {\rm vec}(\tilde{\mathbf{Z}}_{\rm T})-
		(\tilde{\mathbf{A}}_{\rm T}^*\otimes \mathbf{I}_L)\mathbf{z} \Vert_2 <\varepsilon,
	\end{aligned}
\end{equation}
\subsection{2-D CS}

To formulate the channel estimation problem as a 2-D CS problem, we construct the 2-D dictionary using the two 1-D dictionaries $\tilde{\mathbf{A}}_{\rm R}$ and $\tilde{\mathbf{A}}_{\rm T}$. Let $\mathcal{A}$ be the 2-D dictionary and $\mathcal{A}_{i,j}\in\mathbb{C}^{N_Y\times N_X}$ be the $(i,j)$-th atom of $\mathcal{A}$. Futher, 
\begin{equation}
\mathcal{A}_{i,j}=\tilde{\mathbf{a}}_{{\rm R},i}\tilde{\mathbf{a}}_{{\rm T},j}^H,
\end{equation}
 where $\tilde{\mathbf{a}}_{{\rm R},i}$ and $\tilde{\mathbf{a}}_{{\rm T},j}$ are the $i$-th column of $\tilde{\mathbf{A}}_{\rm R}$ and the $j$-th column of $\tilde{\mathbf{A}}_{\rm T}$.
Thus, we derive
\begin{equation}
\tilde{\mathbf{A}}_{\mathrm{R}} \operatorname{diag}(\bm{z}) \tilde{\mathbf{A}}_{\mathrm{T}}^{H} = \sum_{i=1}^{N}{\sum_{j=1}^{N}{z_{i,j}\mathcal{A}_{i,j}}},
	\label{2d_project}
\end{equation}
where $z_{i,j}$ denotes the $(i,j)$-th element of $\bm{z}$.

Based on the 2-D dictionary expression, the 2-D CS based channel estimation problem is given by 
\begin{equation}\label{F2Df}
	\begin{aligned}
		&  \underset{\mathbf{Z}}{\rm arg \ min} \ \Vert \mathbf{Z}\Vert_0\\
		{\rm s.t.} \ &\Vert \tilde{\mathbf{Y}}-\sum_{i=1}^{N}{\sum_{j=1}^{N}{z_{i,j}\mathcal{A}_{i,j}}}\Vert_F \leq \varepsilon.
	\end{aligned}
\end{equation}

\section{Proposed Schemes}
This section will propose effective methods for tackling the channel estimation problem within the three CS frameworks discussed before. To enable a meaningful comparison of the three CS strategies, OMP is employed as the basic algorithm and is expanded to two-stage simultaneous OMP (SOMP) and 2-D OMP for sparse channel estimation.
\subsection{Two-stage SOMP}
SOMP \cite{SOMP}, which is based on the OMP algorithm, is capable of recovering all sparse column vectors in $\mathbf{Z}_{\rm T}$ simultaneously. Obviously, the indexes of the non-zero rows correspond to AoAs. After the matrix $\mathbf{Z}_{\rm T}$ with row-group sparsity is recovered, all zero rows are removed to form a low-dimensional matrix $\tilde{\mathbf{Z}}_{\rm T}$. Then the 1-D CS framework can be used to attain AoDs. Due to the low dimension, the AoD estimation stage is computationally simple.

\subsection{2-D OMP}

\begin{figure*}[hb]
	\hrulefill
	\begin{equation}\label{ls2d}
		\begin{aligned}
			&\mathrm{tr}(\mathbf{Y}_r\mathbf{Y}_r^H)  - 
			\sum_{p^\prime=1}^{p}{z_{i_{p^\prime},j_{p^\prime}}^* \mathrm{tr}(\mathbf{Y}_r\mathcal{A}_{i_{p^\prime},j_{p^\prime}}^H)}  - 
			\sum_{q^\prime=1}^{p}{z_{i_{q^\prime},j_{q^\prime}} \mathrm{tr}(\mathcal{A}_{i_{q^\prime},j_{q^\prime}}\mathbf{Y}_r^H)} +	\sum_{p^\prime=1}^{p}{\sum_{q^\prime=1}^{p}{z_{i_{q^\prime},j_{q^\prime}}z^*_{i_{p^\prime},j_{p^\prime}}\mathrm{tr}(\mathcal{A}_{i_{q^\prime},j_{q^\prime}}\mathcal{A}_{i_{p^\prime},j_{p^\prime}}^H)}}\\
			&\overset{(a)}{=}\left\|\mathbf{Y}_r \right\|_F^2-\sum_{p^\prime=1}^{p}{z_{i_{p^\prime},j_{p^\prime}}^* \left\langle\mathbf{Y}_r, \mathcal{A}_{i_p^\prime,j_p^\prime}\right\rangle}-\sum_{q^\prime=1}^{p}{z_{i_{q^\prime},j_{q^\prime}} \left\langle\mathbf{Y}_r, \mathcal{A}_{i_q^\prime,j_q^\prime}\right\rangle}^*
			+\sum_{p^\prime=1}^{p}{\sum_{q^\prime=1}^{p}{z_{i_{q^\prime},j_{q^\prime}}z_{i_{p^\prime},j_{p^\prime}}^*\left\langle\tilde{\mathbf{a}}_{{\rm T},j_{{p^\prime}}}, \tilde{\mathbf{a}}_{{\rm T},j_{{p^\prime}}}\right\rangle \left\langle\tilde{\mathbf{a}}_{{\rm R},i_{{q^\prime}}}, \tilde{\mathbf{a}}_{{\rm R},i_{{q^\prime}}}\right\rangle}}\\
			&=\left\| \mathbf{Y}_r \right\|_F^2  - 
			\boldsymbol{g}^T \boldsymbol{z}^*-\boldsymbol{g}^H \boldsymbol{z}+ \boldsymbol{z}^H\mathbf{Q} \boldsymbol{z},
		\end{aligned}
\end{equation}
\end{figure*}
Based on the 2-D dictionary described before, we not only extend the 2-D OMP algorithm in the case of real-valued and specialized dictionaries \cite{2domp} to more general applications, but also simplify the 2-D LS algorithm. Algorithm \ref{2domp} illustrates the flow of 2-D OMP.
To put it bluntly, the essential distinction between 1-D and 2-D OMP lies in the matching and LS computing steps. Therefore, we will deduce the matching and LS formulas of 2-D OMP.

\begin{algorithm}[!t] 
	\caption{2-D OMP for MmWave Channel Estimation} 
	\label{2domp}      
	\begin{algorithmic}[1] 
		\footnotesize{
			\REQUIRE {Received signals $\tilde{\mathbf{Y}}$, 2-D dictionary $\mathcal{A}$. }
			\ENSURE {Reconstruction of $\mathbf{Z}$.} 
			
			\STATE{$\textbf{Initialize:}$ $\Lambda=\{(i,j)| \ (1,1),(1,2),\cdots,(N,N)\}$, $\mathcal{I}=\emptyset$, $p=0$ and $\mathbf{Y}_{r,0}=\tilde{\mathbf{Y}}$. 
			}					
			\REPEAT
			\STATE{ \emph{Match 2-D atoms:}  \\ \ \ \ \ \ \ \ \ \ \ $(i_{p},j_{p})\leftarrow {\underset{(i,j)\in\Lambda}{\rm arg \ max}} \ \frac{\left\langle\mathbf{Y}_{r,{p}}, \mathcal{A}_{i,j}\right\rangle}{\left\|\mathcal{A}_{i,j}\right\|_F}$, \\ \ \ \ \ \ \ \ \ \ \ $ 
				\Lambda\leftarrow \Lambda \backslash (i_{p},j_{p})$, $\mathcal{I}\leftarrow\mathcal{I}\cup (i_{p},j_{p})$.}
			
			\STATE{\emph{Comupute 2-D LS:} \\ \ \ \ \ \ \ \ \ \ \
				$\hat{\bm{z}}\leftarrow	\underset{\bm{z}}{\rm arg \ min} \ \Vert \mathbf{Y}_{r,{p}} - \sum_{{p^\prime}=1}^{p}{z_{i_{{p^\prime}},j_{{p^\prime}}}\mathcal{A}_{i_{{p^\prime}},j_{{p^\prime}}}}\Vert_F^2.$
			}
			\STATE{
				\emph{Renew Individual Residual:}\\ \ \ \ \ \ \ \ \ \ \ $\mathbf{Y}_{r,{p}+1}\leftarrow\mathbf{Y}_{r,{p}}-\sum_{{p^\prime}=1}^{p}{\hat{z}_{i_{{p^\prime}},j_{{p^\prime}}}\mathcal{A}_{i_{{p^\prime}},j_{{p^\prime}}}}$.
			}
			
			\STATE{
				$	p\leftarrow p+1$}.
			
			\UNTIL{ $\varepsilon$ is satisfied.}
			
		}
	\end{algorithmic}
\end{algorithm}
\subsubsection{Matching} This is a process of selecting the best atom in each iteration. Let $\mathbf{Y}_r$ be the residual, the projection of $\mathbf{Y}_r$ onto $\mathcal{A}_{i,j}$ is expressed as
\begin{equation}
	\frac{\vert\left\langle\mathbf{Y}_r, \mathcal{A}_{i,j}\right\rangle\vert}{\left\|\mathcal{A}_{i,j}\right\|_F},
\end{equation}
where $\left\langle\mathbf{Y}_r, \mathcal{A}_{i,j}\right\rangle=\tilde{\mathbf{a}}_{{\rm R},i}^H\mathbf{Y}_r\tilde{\mathbf{a}}_{{\rm T},j}$. Then we calculate the projection of $\mathbf{Y}_r$ onto all atoms in the 2-D dictionary $\mathcal{A}$ and select the index corresponding to the largest projection value.

In each iteration, only one atom is selected by maximizing the projection. Let $\{(i_{p^\prime},j_{p^\prime})| \ p^\prime=1,2,\cdots,p\}$ be the entries of the selected atoms after $p$ iterations. 

\subsubsection{Computing 2-D LS} Once the atoms are picked in the matching step, the weights are reassigned in iteration $p$ via the least squares algorithm, which is formulated as
\begin{equation}	
	\underset{\bm{z}}{\rm arg \ min} \ \Vert \mathbf{Y}_r - \sum_{p^\prime=1}^{p}{z_{i_{p^\prime},j_{p^\prime}}\mathcal{A}_{i_{p^\prime},j_{p^\prime}}}\Vert_F^2.
	\label{LS}
\end{equation}	

By using the property of $\Vert \mathbf{V}\Vert_F^2={\rm tr}(\mathbf{V}\mathbf{V}^H)={\rm tr}(\mathbf{V}^H\mathbf{V})$, where $\mathbf{V}$ is an arbitrary complex  matrix,  we can simplify the 2-D LS expression. Hence, the simplified formula of $\Vert \mathbf{Y}_r - \sum_{p^\prime=1}^{p}{z_{i_{p^\prime},j_{p^\prime}}\mathcal{A}_{i_{p^\prime},j_{p^\prime}}}\Vert_F^2$ is given by Eqn. (\ref{ls2d}), shown at the bottom of this page, where 
\begin{equation} 
	\boldsymbol{g} = (\mathrm{tr}(\mathbf{Y}_r \mathbf{D}_{i_1,j_1}^H), \cdots, \mathrm{tr}(\mathbf{Y}_r\mathcal{A}_{i_p,j_p}^H))^T,
\end{equation} 
\begin{equation} 
	\mathbf{Q} = \left(
	\begin{array}{ccc}
		\mathrm{tr}(\mathcal{A}_{i_1,j_1}\mathcal{A}_{i_1,j_1}^H) & \cdots & \mathrm{tr}(\mathcal{A}_{i_1,j_1}\mathcal{A}_{i_p,j_p}^H)\\
		\vdots & \vdots & \vdots\\
		\mathrm{tr}(\mathcal{A}_{i_p,j_p}\mathcal{A}_{i_1,j_1}^H) & \cdots & \mathrm{tr}(\mathcal{A}_{i_p,j_p}\mathcal{A}_{i_p,j_p}^H)
	\end{array}
	\right),
\end{equation}
and (a) holds because of 
\begin{equation}\label{Ar}
	\begin{aligned}
		\mathrm{tr}(\mathbf{Y}_r \mathcal{A}_{i_{p^\prime},j_{p^\prime}}^H) &= \mathrm{tr}(\mathbf{Y}_r \tilde{\mathbf{a}}_{{\rm T},j_{p^\prime}} \tilde{\mathbf{a}}_{{\rm R},i_{p^\prime}}^H)\\
		& = \tilde{\mathbf{a}}_{{\rm R},i_{p^\prime}}^H \mathbf{Y}_r \tilde{\mathbf{a}}_{{\rm T},j_{p^\prime}} \\
		& = \left\langle\mathbf{Y}_r, \mathcal{A}_{i_{p^\prime},j_{p^\prime}}\right\rangle
	\end{aligned}
\end{equation}
and
\begin{equation} 
	\begin{aligned}
		\mathrm{tr}(\mathcal{A}_{i_{{q^\prime}},j_{{q^\prime}}} \mathcal{A}_{i_{{p^\prime}},j_{{p^\prime}}}^H) 
		&= \mathrm{tr}(\boldsymbol{a}_{{\rm R},i_{{q^\prime}}} \tilde{\mathbf{a}}_{{\rm T},j_{{q^\prime}}}^H \tilde{\mathbf{a}}_{{\rm T},j_{{p^\prime}}}\tilde{\mathbf{a}}_{{\rm R},i_{{p^\prime}}}^H)  \\
		&= \left\langle\tilde{\mathbf{a}}_{{\rm T},j_{{p^\prime}}}, \tilde{\mathbf{a}}_{{\rm T},j_{{p^\prime}}}\right\rangle \left\langle\tilde{\mathbf{a}}_{{\rm R},i_{{q^\prime}}}, \tilde{\mathbf{a}}_{{\rm R},i_{{q^\prime}}}\right\rangle.
	\end{aligned}
\end{equation}

Thus, the 2-D LS algorithm is rewritten as 
\begin{equation}	
	\underset{\bm{z}}{\rm arg \ min} \	\left\| \mathbf{Y}_r \right\|_F^2 + \boldsymbol{z}^H\mathbf{Q} \boldsymbol{z} - \boldsymbol{g}^T \boldsymbol{z}^*-
	\boldsymbol{g}^H \boldsymbol{z},
\end{equation}	
which can be solved by 
\begin{equation} 	\frac{\partial{\mathrm{tr}(\mathbf{G})}}{\partial{\boldsymbol{z}}} = \mathbf{Q}^T\boldsymbol{z}^* - \boldsymbol{g}^* = \mathbf{0}
			\Rightarrow \bm{z}=(\mathbf{Q}^H)^{-1}\bm{g},
\end{equation}
where $\mathbf{G}=	\left\| \mathbf{Y}_r \right\|_F^2 + \boldsymbol{z}^H\mathbf{Q} \boldsymbol{z} - \boldsymbol{g}^T \boldsymbol{z}^*-
\boldsymbol{g}^H \boldsymbol{z}$. Note that $\mathbf{Q}$ is a Hermitian matrix, hence $\bm{z}=\mathbf{Q}^{-1}\bm{g}$.

Additionally, we offer a simplified 2-D LS solution which is provided by the following proposition.

\emph{proposition 1:}  The simplified 2-D LS solution is given by
\begin{equation}
\bm{Z}=(\tilde{\mathbf{A}}_{{\rm R}}^H\tilde{\mathbf{A}}_{{\rm R}})^{-1}\tilde{\mathbf{A}}_{{\rm R}}^H\mathbf{Y}\tilde{\mathbf{A}}_{{\rm T}}(\tilde{\mathbf{A}}_{{\rm T}}^H\tilde{\mathbf{A}}_{{\rm T}})^{-1}.
\end{equation}

\emph{proof:} See Appendix \ref{A}.
\subsection{Comparison of different Schemes}
Here, we will make an in-depth analysis of the three schemes, 1-D OMP, two-stage SOMP and 2-D OMP, in recovery performance, computational complexity and memory usage.

\subsubsection{Recovery Performance} Firstly, we analyze the theoretical performances of 1-D and 2-D OMP, shown in the folloing proposition, and put the performance analysis of SOMP in the simulation results.

\emph{proposition 2:} When the OMP algorithm runs, matching 1-D atoms and matching 2-D atoms provide the same results, as do 1-D LS and 2-D LS. Hence, in terms of the signal recovery performance, 1-D and 2-D OMP are equivalent. 

\emph{proof:} See Appendix \ref{B}.

\subsubsection{Computational Complexity}

The complexity discussed next refers to the computational complexity in an iteration. Whether 1-D OMP, two-stage SOMP or 2-D OMP, their complexity is mostly due to the matching step. This is because LS is calculated using a very small number of iterations, implying that its computational complexity is negligible in comparison to the matching step. 

Matching 1-D atoms requires $(N^2\times N_XN_Y)\times(N_XN_Y\times 1)$ matrix-vector multiplication to caculate the projection of $\mathbf{y}_r$ onto all atoms, i.e., $\tilde{\mathbf{A}}^H\mathbf{y}_r$, where $\mathbf{y}_r$ is the vectorization of $\mathbf{Y}_r$. This incurs a complexity of $\mathcal{O}(N^2N_XN_Y)$. Furthermore, 
The complexity of the 2-D matching process is mainly dominated by $(N\times N_Y)\times(N_Y\times N_X)$ matrix-matrix multiplication and $(N\times N_X)\times (N_X\times N)$ matrix-matrix multiplication. Since $N_Y<N$, this results in a complexity of $\mathcal{O}(N^2N_X)$. For two-stage SOMP, the matching step of the first stage dominates the complexity, which is $\mathcal{O}(N^2N_Y)$.

In addition, we discuss the complexity of 1-D LS, 2-D LS and simplified 2-D LS directly for channel estimation. To successfully use the LS algorithm for channel estimation, $N^2=N_XN_Y$ is set. According to the inverse operation, 1-D and 2-D LS incur a complexity of $\mathcal{O}(N_X^3N_Y^3)$ while simplified 2-D LS leads to a low complexity of $\mathcal{O}({\rm max}(N^3_X,N^3_Y))$.





\subsubsection{Memory Usage} The key difference of memory usage between the three schemes is that 1-D OMP requires the complexity of $\mathcal{O}(N^2N_XN_Y)$ to store  $\tilde{\mathbf{A}}$, but two-stage SOMP and 2-D OMP only need the complexity of $\mathcal{O}(N{\rm max}(N_X,N_Y))$ to store $\tilde{\mathbf{A}}_{\rm T}$ and $\tilde{\mathbf{A}}_{\rm R}$.

\section{Simulation Results}
In this section, numerical simulations are carried out to demonstrate the effectiveness of the proposed schemes, where the 1-D OMP, 1-D LS and SBL methods are benchmarks. 
 The simulation parameters are described as follows. $N_{RF}=4$, $N_t=N_r=64$, $L=3$,  $d=\lambda/2$, AoAs and AoDs are chosen randomly in $[0,\pi]$. The signal to noise (SNR) is defined as SNR=$\sigma_p^2/\sigma_n^2$.

\begin{figure}[htbp]\label{nmse}
	
	\centering
	
	\includegraphics[width=7cm,height=5.2cm]{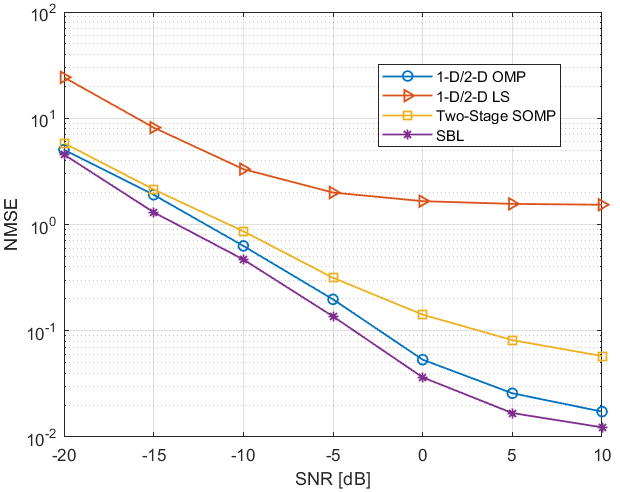}
	\centering
	\caption{NMSE of different methods versus SNR.  }
\end{figure}
\figurename{2} exhibits the normalized mean square error (NMSE) of different schemes in the cases when SNR ranges from -20 dB to 10 dB. In this simulation paramater setting, $N=32$ and $N_X=N_Y=12$. In particular, $N=12$ is set for 1-D and simplified 2-D LS to ensure their successful use. This figure illustrates that two-stage SOMP has a worse performance than 1-D and 2-D OMP. Additionally, SBL exceeds OMP since the former is a global optimization, but it will lead to very large computational complexity. Finally, owing to the incredibly restricted number of angle grid points in 1-D and simplified 2-D LS, they perform poorly.

\begin{figure}[htbp]

	\centering
	\includegraphics[width=7cm,height=5.2cm]{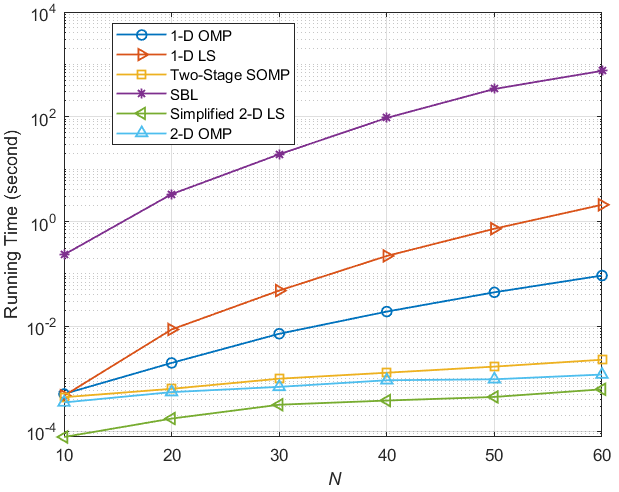}
	\centering
	\caption{Running time comparison of different methods. }
\end{figure}
In order to reveal the superiority of the proposed schemes in computational complexity, \figurename{3} depicts the running time comparison versus $N$ of different schemes, where $N=N_X=N_Y$ and SNR=10 dB. We can observe that OMP based methods have a lower computational complexity than SBL, whereas two-stage SOMP and 2-D OMP have a comparable computational complexity but are about $N$ times faster than 1-D OMP. Furthermore, simplified 2-D LS decreases the complexity of 1-D LS to the point that it becomes the fastest method.

\section{Conclusions}\label{Con}
In the majority of channel estimation research, the conventional 1-D CS framework is employed with very high computational complexity. For efficient channel estimation, we suggest two low-complexity frameworks:  one separates the estimation of AoAs and AoDs into two stages, while the other recovers the 2-D sparse matrix corresponding to AoAs and AoDs. Furthermore, comparing different approaches' recovery performance and computational complexity is made to demonstrate the strengths of our proposed strategies. In our future work, we tend to exploit off-grid theory and sparsity structures, such as joint sparsity, to extend our low-complexty concept to multi-user, frequency selective and beam squint effect based channel estimation.

\begin{appendices}
	\section{  }\label{A}

 It is well established that 1-D LS is caculated by $\hat{\bm{z}}=(\tilde{\mathbf{A}}^H\tilde{\mathbf{A}})^{-1}\tilde{\mathbf{A}}^H\mathbf{y}$, where
\begin{equation}\label{1DLS}
	\begin{aligned}
		\tilde{\mathbf{A}}^H\tilde{\mathbf{A}}=&(\tilde{\mathbf{A}}_{\rm{T}}^*\otimes\tilde{\mathbf{A}}_{\rm R})^H(\tilde{\mathbf{A}}_{\rm{T}}^*\otimes\tilde{\mathbf{A}}_{\rm R})\\
		=&(	\tilde{\mathbf{A}}_{\rm T}^T	\tilde{\mathbf{A}}_{\rm T}^*)\otimes(\tilde{\mathbf{A}}_{\rm R}^H\tilde{\mathbf{A}}_{\rm R}).
	\end{aligned}
\end{equation}
Therefore, the solution of simplified 2-D LS can be deduced by that of 1-D LS. In line with Eqn. (\ref{1DLS}), we have
\begin{equation}
	\begin{aligned}
	\bm{z}=&((	\tilde{\mathbf{A}}_{{\rm T}}^T	\tilde{\mathbf{A}}_{{\rm T}}^*)\otimes(\tilde{\mathbf{A}}_{{\rm R}}^H\tilde{\mathbf{A}}_{{\rm R}}))^{-1}(\tilde{\mathbf{A}}_{{\rm T}}^*\otimes\tilde{\mathbf{A}}_{{\rm R}})^H\mathbf{y}\\
	=&(	(\tilde{\mathbf{A}}_{{\rm T}}^T	\tilde{\mathbf{A}}_{{\rm T}}^*)^{-1}\tilde{\mathbf{A}}_{{\rm T}}^T)\otimes((\tilde{\mathbf{A}}_{{\rm R}}^H\tilde{\mathbf{A}}_{{\rm R}})^{-1}\tilde{\mathbf{A}}_{{\rm R}}^H)\mathbf{y},
	\end{aligned}
\end{equation}
hence,
\begin{equation}
	\bm{Z}={\rm devec}(\bm{z})=(\tilde{\mathbf{A}}_{{\rm R}}^H\tilde{\mathbf{A}}_{{\rm R}})^{-1}\tilde{\mathbf{A}}_{{\rm R}}^H\mathbf{Y}\tilde{\mathbf{A}}_{{\rm T}}(\tilde{\mathbf{A}}_{{\rm T}}^H\tilde{\mathbf{A}}_{{\rm T}})^{-1},
\end{equation}
where ${\rm devec}(\cdot)$ denotes the devectorization operation.
	\section{  }\label{B}
	To demonstrate that 1-D and 2-D OMP have the same performance, we need simply to establish that 1-D and 2-D matching are equivalent, since 1-D and 2-D LS solutions are identical and can be derived as follows.
	
	Suppose that $\Omega$ is the selected support in the $p$-th iteration of 1-D OMP. Thus, the $(i,j)$-th element of $\tilde{\mathbf{A}}_\Omega^H\tilde{\mathbf{A}}_\Omega$ equals the $(i,j)$-th element of $\mathbf{Q}$, i.e., $[\tilde{\mathbf{A}}_\Omega^H\tilde{\mathbf{A}}_\Omega]_{i,j}=\left\langle\tilde{\mathbf{a}}_{{\rm T},j}, \tilde{\mathbf{a}}_{{\rm T},j}\right\rangle \left\langle\tilde{\mathbf{a}}_{{\rm R},i}, \tilde{\mathbf{a}}_{{\rm R},i}\right\rangle$. For instance, suppose only one atom in $\Omega$ is $\tilde{\mathbf{a}}_{N(j-1)+i}$, then $\tilde{\mathbf{A}}_\Omega^H\tilde{\mathbf{A}}_\Omega=(\tilde{\mathbf{a}}^T_{{\rm T},j}\tilde{\mathbf{a}}^*_{{\rm T},j})\otimes(\tilde{\mathbf{a}}^H_{{\rm R},i}\tilde{\mathbf{a}}_{{\rm R},i})=\left\langle\tilde{\mathbf{a}}_{{\rm T},j}, \tilde{\mathbf{a}}_{{\rm T},j}\right\rangle \left\langle\tilde{\mathbf{a}}_{{\rm R},i}, \tilde{\mathbf{a}}_{{\rm R},i}\right\rangle$.
	
	Besides, $\bm{g}=\tilde{\mathbf{A}}_\Omega^H\mathbf{y}_r$ can be easily deduced from Eqn (\ref{Ar}). Then we have
	$\bm{z}=(\tilde{\mathbf{A}}_\Omega^H\tilde{\mathbf{A}}_\Omega)^{-1}\tilde{\mathbf{A}}_\Omega^H\mathbf{y}_r=\mathbf{Q}^{-1}\bm{g}$,
	which follows that the solutions of 1-D and 2-D LS are identical.
	
	Mathcing 1-D atoms needs to compute the projection of the 1-D residual $\mathbf{y}_r$ onto all 1-D atoms in $\tilde{\mathbf{A}}$, where the $(N(j-1)+i)$-th atom of $\tilde{\mathbf{A}}$ is the vectorization of $\mathcal{A}_{i,j}$ because
	\begin{equation}
		{\rm vec}(\mathcal{A}_{i,j})={\rm vec}(\tilde{\mathbf{a}}_{{\rm R},i}\tilde{\mathbf{a}}_{{\rm T},j}^H)
		=\tilde{\mathbf{a}}_{{\rm T},j}^*\otimes\tilde{\mathbf{a}}_{{\rm R},i}
		=\tilde{\mathbf{a}}_{N(j-1)+i},
	\end{equation}
	where $\tilde{\mathbf{a}}_{N(j-1)+i}$ is the $(N(j-1)+i)$-th atom of $\tilde{\mathbf{A}}$.
	
	Thus, we have $\Vert \mathcal{A}_{i,j}\Vert_F=\Vert\tilde{\mathbf{a}}_{N(j-1)+i}\Vert_2$. Then the projection of $\mathbf{y}_r$ onto the atom $\tilde{\mathbf{a}}_{N(j-1)+i}$ satisfies
	\begin{equation}
		\begin{aligned}
			\frac{ \left\langle\mathbf{y}_r,\tilde{\mathbf{a}}_{N(j-1)+i}\right\rangle}{\Vert\tilde{\mathbf{a}}_{N(j-1)+i}\Vert_2} =&\frac{\sum_{m=1}^{N_Y}\sum_{n=1}^{N_X}(a_{{\rm R},i}^m)^*a_{{\rm T},j}^n y_r^{N(m-1)+n}}{\Vert\tilde{\mathbf{a}}_{N(j-1)+i}\Vert_2}\\
			=& \frac{\tilde{\mathbf{a}}_{{\rm R},i}^H\mathbf{Y}_r\tilde{\mathbf{a}}_{{\rm T},j}}{\Vert\tilde{\mathbf{a}}_{N(j-1)+i}\Vert_2}
			=\frac{\left\langle\mathbf{Y}_r, \mathcal{A}_{i,j}\right\rangle}{\Vert \mathcal{A}_{i,j}\Vert_F},
		\end{aligned}
	\end{equation}
	where $a^m_{{\rm R},i}$ denotes the $m$-th element of the atom $\tilde{\mathbf{a}}_{{\rm R},i}$, similarly, $a^n_{{\rm T},j}$ is the $n$-th element of the atom $\tilde{\mathbf{a}}_{{\rm T},j}$ and $y_r^{N(m-1)+n}$ represents the $(N(m-1)+n)$-th element of $\mathbf{y}_r$. This implies that the atom selection of the two algorithms is the same in each iteration. 

\end{appendices}

\bibliographystyle{IEEEtran}
\bibliography{reference.bib}

\vspace{12pt}

\end{document}